# The Hopping Mechanism of the Hydrated Excess Proton and Its Contribution to Proton Diffusion in Water

Christopher Arntsen,[a,1] Chen Chen,[b,1] Paul B. Calio,[c] Chenghan Li,[c] and Gregory A. Voth[c,*]

[a]Department of Chemistry, Youngstown State University, Youngstown, Ohio, 44505 USA
[b]Department of Mechanical and Nuclear Engineering, Pennsylvania State University, University Park, Pennsylvania, 16802 USA
[c]Department of Chemistry, Chicago Center for Theoretical Chemistry, James Franck Institute, and Institute for Biophysical Dynamics, The University of Chicago, Chicago, Illinois, 60637 USA

**Abstract**

In this work a series of analyses are performed on *ab initio* molecular dynamics (AIMD) simulations of a hydrated excess proton in water to quantify the relative occurrence of concerted hopping events and "rattling" events, and thus to further elucidate the hopping mechanism of proton transport in water. Contrary to results reported in certain earlier papers, the new analysis finds that concerted hopping events do occur in all simulations, but that the majority of events are the product of proton rattling, where the excess proton will rattle between two or more waters. The results are consistent with the proposed "special-pair dance" model of the hydrated excess proton, wherein the acceptor water molecule for the proton transfer will quickly change (resonate between three equivalent special pairs), until a decisive proton hop occurs. To remove the misleading effect of simple rattling, a filter was applied to the trajectory such that hopping events that were followed by back hops to the original water are not counted. A steep reduction in the number of multiple hopping events is found when the filter is applied, suggesting that many multiple hopping events that occur in the unfiltered trajectory are largely the product of rattling, contrary to prior suggestions. Comparing the continuous correlation function of the filtered and unfiltered trajectories, we find agreement with experimental values for the proton hopping time and Eigen-Zundel interconversion time, respectively.

---

[*] To whom all correspondence should be addressed. E-mail: gavoth@uchicago.edu
[1] These authors contributed equally to this work.

**I. Introduction**

Proton transport is an important chemical process fundamental to a number of systems within the fields of biology, chemistry, materials science, and engineering. Such systems include the proton transport, coupling, and pumping mechanisms within proteins,[1-4] as well as proton exchange membranes (a common fuel cell material),[5-7] the efficient functionalities of which rely on facile and rapid proton transport. Unlike other cations, the transport of a proton in aqueous media relies on two distinct mechanisms: vehicular transport and Grotthuss shuttling, i.e., proton hopping from hydronium ions to neighboring water molecules.[8-18] The ability of a hydrated excess proton to hop between neighboring waters allows the charge center to diffuse without substantial displacement of nuclear coordinates and significantly increases the excess proton charge center self-diffusion constant.

In order to fully understand the mechanism of proton transport in experimental systems, a comprehensive and physically rigorous theoretical framework is necessary. Given the unique character of proton transport – i.e., involving chemical bond rearrangement – any such description must necessarily incorporate proton transfer from hydronium ions to neighboring water molecules. Since bond breaking and formation is fundamentally a quantum mechanical process involving the rearrangement of electrons, one natural choice is to employ *ab initio* molecular dynamics (AIMD) simulation methods to investigate the proton hopping mechanism. Despite years of effort from a number of research groups, there is a still considerable debate regarding the nature of the proton transport mechanism, especially its hopping component. The importance of multiple, concerted proton hops as a driver of enhanced proton charge diffusion in particular has remained contentious.

Much of the debate is centered around the true nature of a hydrated excess proton; that is, whether it is most accurately described as existing as an Eigen cation, $H_9O_4^+$,[19] or a Zundel cation, $H_5O_2^+$,[20] although these are clearly just limiting cases.[17] Dominance of single proton hopping involves the conversion of the (distorted) Eigen cation into a transient Zundel cation and to another Eigen structure, resulting in transfer of a proton from one $H_3O^+$ core structure to another.[21-23] By contrast, some simulations have suggested direct conversion of one Zundel ion into another, resulting in a double hop.[10, 24-26] However, the early AIMD work by Marx et al.,[13] and the multistate empirical valence bond (MS-EVB) model of Schmitt and Voth,[12] suggested that both the Eigen and Zundel are only limiting structures, and the excess proton is actually continually in a state of flux between them. This latter work found that proton hopping was dominated by single proton transfer events from one hydronium core to another, and modulated by hydrogen bonding rearrangements in the solvating water molecules.

Much of the debate around the proton hopping mechanism has focused on the importance of the first solvation shell of the hydronium ion. Early work suggested that due to the reduced coordination of a hydronium ion relative to that of a water molecule (three in hydronium vs. four in water),[10, 27] proton transfer is driven by cleavage of one of the hydrogen bonds (HB) donated to the proton acceptor and by the subsequent formation of an additional hydrogen bond to the proton donor. In fact, Voth, Agmon, Tuckerman and co-workers found that in hopping events which resulted in charge displacement – i.e., neither rattling nor a special pair dynamics – involved the *concerted* cleavage and formation of these hydrogen bonds.[21, 28] In this mechanism, the cleavage of a HB donated to the proton acceptor happens first, followed by formation of a HB donated to the donor (often within 50 fs); this particular order does not occur in every hopping instance. On the other hand, it has been shown that the average coordination of bulk water is around 3.9, whereas that of water in the first solvation shell of the hydrated proton is around 3.6.[29-30] This suggests that cleavage of a HB donated to the acceptor water molecule is not the rate limiting step. Subsequent work by Tse et al.[31] showed that by examining the $O^*$-$H_w$ radial distribution function (where the asterisk denotes atoms belonging to core hydronium-like ions), proton transfer was facilitated by the presence of a fourth water donating a hydrogen bond to the top of the oxygen atom of the hydronium ion. Consistent with the results of Tuckerman and co-workers, it was found that charge diffusion was enhanced in regions of the trajectory in which this water was present, resulting in burst and rest periods characteristic of AIMD.

In an extensive analysis of both MS-EVB and AIMD trajectories, Voth, Agmon, and co-workers[22] found that the proton transport process involves even more steps than the formation and cleavage of two hydrogen bonds. It was found that the first solvation shell water nearest to the hydronium ion rapidly shifts; which is termed the "special pair dance", and in this process, the hydronium is "choosing" to which water to donate the proton. Earlier work by Lapid et al.[21] found that it is the collective motion of the first two solvation shells in determining which of the waters to which the excess proton is ultimately donated. This analysis involves comparing both the relative bond lengths of hydrogen bonds between first and second solvation shell waters, as well as formation and cleavage of hydrogen bonds between these layers. Therefore, rather than be dictated by just a few atoms or molecules, the proton transfer process involves upwards of twenty waters in an extended hydrated proton complex. This result, coupled with the duration of the special pair in time, indicates that single hops should dominate proton hopping events, as the extensive solvation structure simply cannot react fast enough to accommodate multiple hops.

In a recent paper, however, Hassanali and co-workers argued that the formation of water wires is abundant in bulk water, resulting in periods with bursts of proton hopping activity.[32] This has been observed in several AIMD studies, and the authors of Ref. [32] argue that proton transfer occurs in concerted hopping events over several waters, and is dominated by double hops. More recently, Chen et al. observed similar behavior.[33] These authors argued in fact that AIMD simulations of an excess proton in water are overwhelmingly dominated by correlated double hopping events, leading to the observed more rapid protonic charge defect diffusion.

The arguments in Ref. [33] that double hops dominate proton transfer contradicts many of the results seen previously. In this work, we have therefore re-evaluated a set of AIMD simulations in an attempt to further clarify the role of single hops and concerted double hops in the proton transport. We use several analyses to count and assess the abundance of multiple hopping events relative to the total number of events. Our extensive analysis finds that single hops are the prevalent proton hopping mechanism for the hydrated excess proton, in contrast to the multiple hopping behavior suggested in earlier work.

The organization of this paper is as follows: in Section II we briefly discuss the simulation details; in Section III we discuss the analysis methods employed, as well as the resulting data; and in Section IV we present a collective discussion of the results and provide our conclusions on the presence and importance of multiple hops.

**II. Simulation Details**

The results presented below are based on duplicate sets of the four simulation setups: AIMD at 300 K, AIMD at 330 K, and EDS-AIMD (OO) at 300 K, and EDS-AIMD (OH) at 300 K. EDS-AIMD is an *ab initio* method where the simulation has been minimally biased to match experimental data, hence the name EDS: experimentally directed simulation.[34] In EDS-AIMD (OO), a bias is applied such that the $O_w$-$O_w$ RDF in a pure water system is reproduced; it has also been found that this bias does not distort the solvation structure of a hydronium ion.[34-35] In EDS-AIMD (OH),[36] a bias is applied such that the $O_w$-$H_W$ RDF reproduces that of classical MB-pol water,[37-40] and forces were applied continuously between any O-H pairs;[41] details of this methodology will be presented in a following publication. Inclusion of AIMD simulations at 330 K was motivated by the fact that all the simulations in Ref. [33] were run at that elevated temperature. Each of the calculations used the BLYP exchange-correlation functional,[42-43] with the Grimme dispersion correction.[44-45] Each simulation used a triple-zeta basis set to describe the valence electrons, and a Goedecker-Teter-Hutter pseudopotential. The simulations were equilibrated for at least 20 ps in the constant NVT ensemble using the particular AIMD method and temperature of the target simulation. Finally, duplicate production runs of the systems in the constant NVE ensemble were run for 80 ps. The timestep used for the simulations was 0.5 fs. All AIMD simulation were run using the CP2K software package,[46] and the EDS-AIMD simulations used the EDS extension within PLUMED.[47-48]

## III. Results

### A. Hopping

*i. Hopping Counting and Filtering*

There are several ways in which one could count the number of single and multiple hopping events, and we thus present the hopping behavior of a hydrated excess proton in water using three counting methods as described below. In accordance with the work in Ref. [33], we have chosen the relevant time scale of 500 fs. To eliminate double counting, we scan the trajectory, looking to maximize the number of hops in a given block of length 500 fs. Again, in keeping with the work in Ref. [33], we tabulate the number of hops in segment lengths of 10 ps. We note that in each counting method, if the proton hops from water A to water B, then back to water A, this is not considered a hop. It is instead a "rattle".

In Method 1, we scan the trajectory for the largest number of hops within 500 fs, and if we find an $n$ hopping event in that 500 fs range, we do not scan that region of the trajectory again. We essentially tile the trajectory with blocks 500 fs long in which there are the most number of many hopping events. That is, we scan for $n = 10, n = 9, \ldots, n = 1$.

Method 2 is very similar to Method 1, but segment lengths of 500 fs are not required. For example, suppose that at the end of a 500 fs block, there were 3 hops. Now suppose that the third hop occurred at, say, 350 fs into that 500 fs block. Then the "tile" pertaining to the $n = 3$ event is just the first 350 fs. Thus, this region cannot contribute to any other hopping events, but the latter 150 fs is free to contribute to another hopping event.

Method 3 is a slight modification which highlights most favorably multiple hopping events. Again, consider a 500 fs block in which the final number of hops at the end of the block is $n = 3$. Now suppose that in this block, there was a point at which $n = 4$ before some subsequent back-hop, and say this hop which resulted in $n = 4$, albeit fleetingly, occurred 250 fs into the block. We now tile the first 250 fs with $n = 4$, and leave the remaining 250 fs open to contribute to other hopping events. We recognize that what seem like favorable multiple hopping events may be nullified by a subsequent back-hop, but Method 3 seeks to find the most favorable multiple hopping situation.

The results for the above analysis methods are shown in Fig. 1. One can argue the number of single hops in each of the analysis methods is overestimated as a result of the tiling technique used. For example, consider two tiles that are nearly adjacent, with a narrow portion of the trajectory separating the two. This small portion of the trajectory is naturally going to be limited in the number of hops it contains as a result of its limited size. We recognize this complicates the description of single hops, but as we generally disagree with the conclusion that multiple hops dominate, we set out to be as generous to multiple hopping events as possible.

The results show several interesting trends. First, as expected, there is an increase in the number of multiple hops as one moves from Method 1 to Method 2 to Method 3. This is clearly a result of a more generous tiling scheme, allowing for more multiple hopping events and more events with many hops. Another notable feature is the substantial increase in multiple hopping events when the temperature of AIMD simulations is increased from 300 K to 330 K. This is somewhat unsurprising as the free energy barrier for proton transfer of AIMD at 330 K is somewhat lower than that of AIMD at 300 K (Fig. S1 of the Supporting Information), and the increased thermal energy should further make the transition from one water to another easier. It is also interesting that EDS-AIMD, run at 300 K, produces an intermediate number of multiple hops.

Again, given the slightly lower proton transfer barrier, one could expect a slightly larger number of hops.

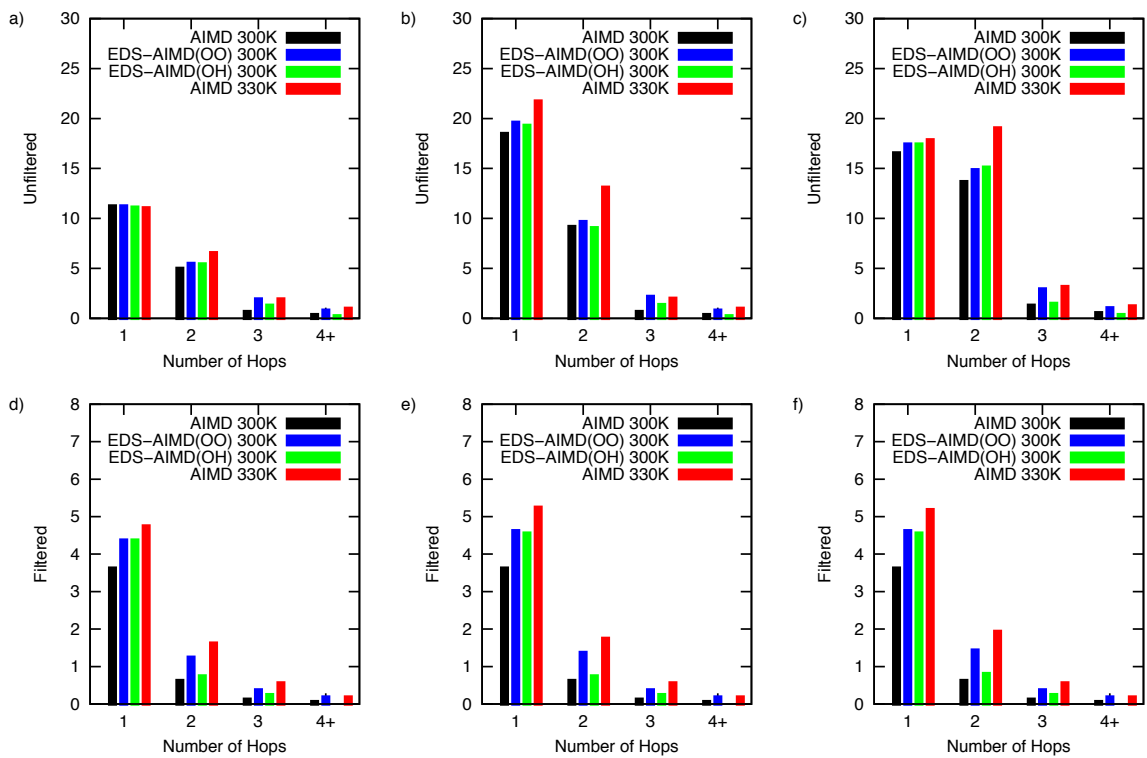

**Figure 1:** Number of occurrences of different hopping method for the unfiltered (a-c) and filtered (d-f) trajectories. The numbers reported are averages of 10 ps fragments of the trajectory.

AIMD water with various functionals (including BLYP) is known to be overstructured,[49-50] and this can sometimes result in more rapid proton transfer, as neighboring waters are especially ready and "sticky in place" to receive an excess proton. On the other hand, it has been shown that cleavage of hydrogen bonds in the first and even second solvation shells of the Eigen complex directly impact the proton transport mechanism. Thus, overstructuring may prevent waters from adopting a configuration capable of receiving a transferred proton. Given that EDS is designed explicitly to reduce the structure of water, and elevating the ionic temperature of a simulation from 300 to 330 K would naturally have such an effect, it is possible that the disruption of the water hydrogen bond network in fact makes back-transfer more difficult.

One of the limitations of the above analysis is that it can be corrupted by back-hopping (i.e., situations in which a proton hops from water A to water B, and then back to water A). We note that back-hopping is quite common in AIMD simulations, and results in proton rattling between two waters. Consider an example in which the proton back-hops before immediately hopping forward. If we label the waters alphabetically in terms of forward hopping, this would look something like A → B → A → C, where A → B is the initial forward hop, B → A is the back hop, and then A → C is another forward hop (note that waters B and C are equivalent in the forward hopping chain). The above analysis could pick out the sequence B → A → C as a double hop, even though a back hop is required to make the double hop possible. Our analysis could also pick out hopping events that are immediately followed by a back hop, again overemphasizing high *n* events.

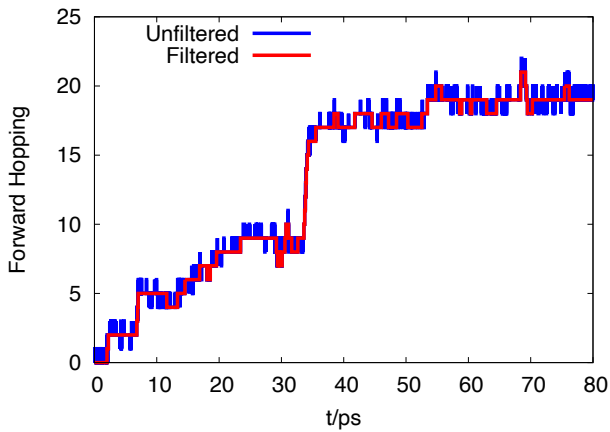

**Figure 2:** Forward hopping index as a function of time for the unfiltered (blue) and filtered (red) trajectories.

For the purpose of this analysis, we have applied a filter to the trajectory that mitigates the effect of rattling by smoothing out such events. If a proton hops from A to B, and then back to A, we consider the proton to have remained on A the entire duration. The only exception to this is if the proton remains on B for 500 fs or more, in which case we consider B to be a stable hydronium, rather than just a fleeting hydronium. To illustrate the effect of this filter, we show an example of the filtered and unfiltered forward hopping of an AIMD simulation at 300 K in Fig. 2 (the effect of the filter is the same for all the simulations, so we show just one example; analysis of all trajectories can be found in Fig. S2). Comparing the forward hopping of the filtered and unfiltered trajectories, we see the clear elimination of a large proportion of hops, and within the context of concerted hopping, a majority of concerted hops are eliminated. Notice that even in the filtered trajectory there is some apparent rattling – our filter will only eliminate first order rattling; that is, if the proton hopping goes forward and back as in A → B → C → B → A – a term we will call the "slingshot" effect – the filter will only trim the trajectory of the hop to C, resulting in A → B → A.

We ran the above hop counting techniques on the filtered trajectories, and found a dramatic reduction in the number single and multiple hopping events in all three methods. This is unsurprising based on the fact that we have eliminated the "slingshot" effect mentioned above, and therefore trimmed the maximum number of effective hops. However, it is somewhat surprising how *drastically* the number of hops is reduced. Comparing the relative numbers of single hops to multiple hops again suggests the dominance in single hopping events. We again concede the possibility of overcounting single hops as a result of "bumping" against a tile. Nonetheless, if the prevalence of multiple hops seen in the unfiltered trajectory were genuinely concerted, and not some artifact of the slingshot effect, then we would expect to see most of the multiple hopping events maintained in the filtered trajectory. This is not the case, based both on the results of the analysis and simple visual inspection of the forward hopping plot shown in Fig. 2.

As in the above analysis, we find that after applying the filter, there is still a strong temperature dependence on the number of concerted hops: AIMD simulations at 330 K have a much higher propensity for double hops (and higher order hopping events) than AIMD at 300 K. This suggests one must exercise caution when interpreting results of AIMD simulations at higher temperatures. We also see an intermediate number of double hops in the EDS-AIMD simulations between AIMD at 300 K and AIMD at 330 K.

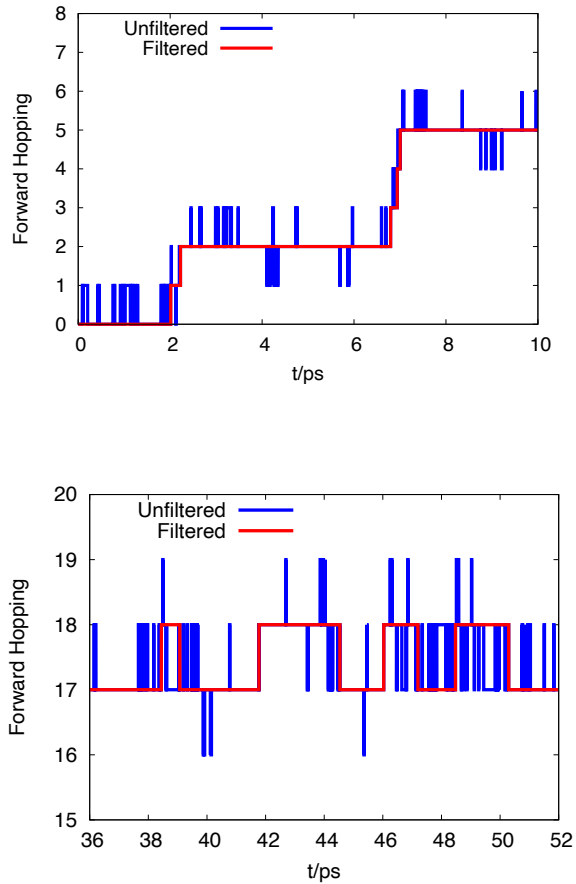

**Figure 3:** Segments of the filtered and unfiltered trajectories of AIMD at 300 K showing the effect of slingshotting.

Another striking feature of this data is that the majority of hopping events are single hopping events. That is, when one eliminates rattling and slingshotting, one eliminates what we argue had been mistakenly categorized as concerted hops. To illustrate this point, consider the zoomed in region of the AIMD forward hopping plot shown in Fig. 3, showing the forward hopping in the first 10 ps of the simulation. Consider the hopping event at around 2 ps. The filtered trajectory in red shows an unambiguous double hopping event. The unfiltered trajectory shows some rattling, a double hop, and then some more rattling. There is a point at which there is a net *n* = 3 event, but this is mitigated by a near-immediate back hop. It is clearly nonsensical to assign this to a triple hopping event, considering that the third hop is so quickly nullified. One advantage of using the filtered trajectory is that it allows easy visual inspection and verification. Our analysis also relates a reasonable number of hopping events relative to the forward hopping. In the example trajectory shown in Fig. 3, there are 1037 total hops, the net forward hopping is only 20. It therefore seems unreasonable to mistakenly consider the system to be undergoing constant concerted hopping.

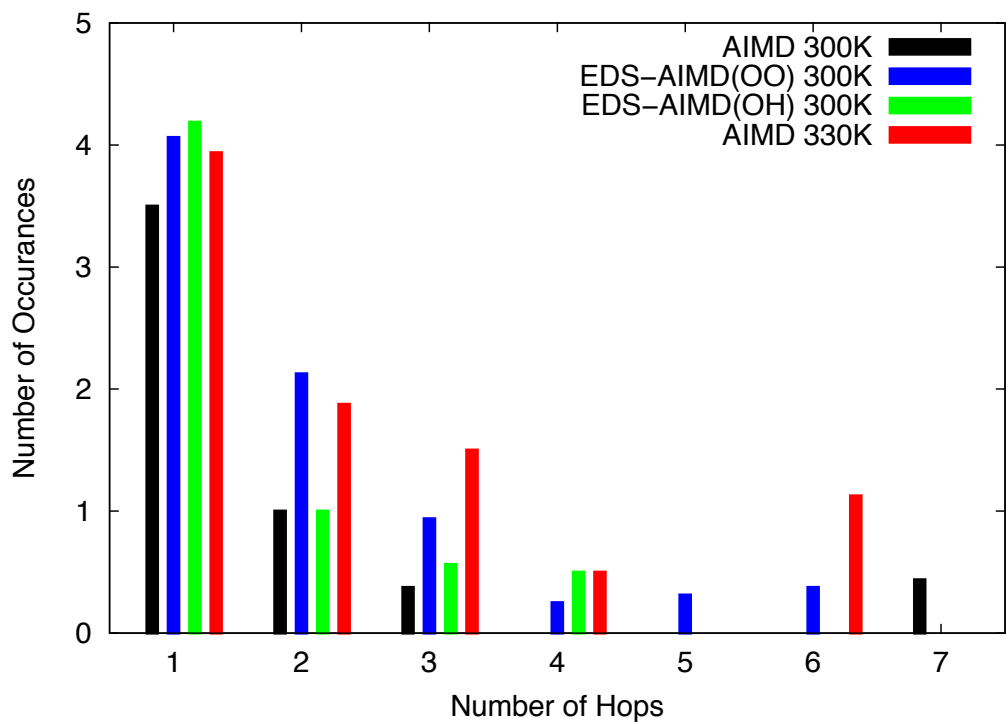

**Figure 4:** Analysis of hopping event using the method described in Ref [33]. Due to the fact that there are so few segments in which a proton stays on a single water for 0.5 ps, results shown are only for the filtered trajectory. Number of Occurrences are averaged over 10 ps segments of the trajectory.

We also calculated the forward hopping that occurs in the 0.5 ps immediately following a hopping event; that is, when a hopping event occurs, what is the net relative forward hopping 0.5 ps later. Of the 1037 total hopping occurrences, 43% of the hopping occurrences had the proton return to the same position in the forward hopping chain; 21% of the hopping occurrences had the proton one position forward; and 4% of the hopping occurrences the proton was two positions forward. Only 6% occurrences had the proton two or more positions forward after 0.5 ps. (The other occurrences are attributed to net negative forward hopping after a hopping event.) When we run the same analysis on the individual simulation with the fastest diffusion, AIMD at 330 K having excess proton self-diffusion constant of 2.07 Å$^2$/ps, we do find more forward hopping and concerted hopping, but not in such a way as to suggest a dominance of concerting hopping. In this particular run at 330 K, there were 1435 total hopping events; in 34% of those, the proton was at the same forward hopping position 0.5 ps later; in 26% the proton was one position ahead; in 11% the proton was two positions ahead; and in 14% the proton was two or more positions ahead. That is, the proton only moves forward in a concerted way around 14% of the time.

We will not discuss each individual hopping event in the trajectory shown in Fig. 3, but it is instructive to elaborate on another specific region in the trajectory. Consider the region of the forward hopping plot of the AIMD trajectory at 300 K between 36 and 52 ps. In the unfiltered trajectory, there are a number of clear hopping events, and even a few hopping events that could be construed as double hops. However, it is clear, looking at the forward hopping of both the filtered and unfiltered trajectories, that essentially no forward hopping is occurring, and therefore it would be unreasonable to assign any of the action as a multiple hopping event. Indeed, incessant rattling is most prominent in AIMD simulations at 300 K, but such periods of apparent activity that are accompanied by no forward hopping do occur in simulations run with AIMD at 330 K and EDS at 300 K, though in AIMD at 330 K extensive periods of inactivity are rare.

To more directly compare with the analysis in Ref. [33], we have also implemented the following scheme for counting the number of hops. The number of hops in a concerted hopping event is incremented as the proton hops to a new water; if the proton back hops, the previous hop is nullified; a concerted hopping event is concluded if the proton remains on a water for 0.5 ps. This protocol is taken from Ref. [33] As can be seen from the unfiltered forward hopping plot in Fig. 2, the excess proton essentially never sits on a water for 0.5 ps (and we make no effort to analyze the unfiltered trajectory in this way). We therefore run our analysis on the filtered trajectory. We note that Ref. [33] makes no mention of filtering in the specific way that we do, but the authors do state they eliminate counting of rattling events by considering a proton that has returned to its original water within 0.5 ps not to have hopped. We reiterate that protons so infrequently stay on a single water for 0.5 ps, and that rattling precedes nearly every hopping event (though not exactly all), thus necessitating the use of the filtered trajectory.

To present our data as consistently with Ref. [33], we scale the number of *n* hopping events by *n*: if there are 10 double hops, we plot $2 \times 10 = 20$ (we note this is what was done in Ref. [33]). The results of this analysis are shown in Fig. 4. By using the filtered trajectory, we again find a dramatic reduction in the number of multiple hops, and by imposing the condition that the proton must remain on the same water for 0.5 ps to conclude a single or multiple hopping event, there are simply fewer total events. We note that we did not plot $n = 0$ events, in which a proton sits on a single water for 0.5 ps; this is a result of filtering out rattling events, as such a scenario is virtually never the case in an unfiltered trajectory.

In accordance with the abundance of single hopping events is the relative lack of multiple hopping events in comparison to Ref. [33], despite having implemented a protocol very similar to what those authors report, though admittedly our filter may have disrupted an apples to apples comparison. Multiple hopping events are most significant in AIMD at 330 K, with the largest number of multiple hopping events coming in the form of double and triple hops. We do see a similar number of double hops in EDS-AIMD (OO) at 300 K, and an appreciable number of higher order multiple hops, but fewer than in AIMD at 330 K. We find few multiple hopping events in AIMD at 300 K, though interestingly there is one $n = 7$ hopping event. We note again that AIMD with the BLYP functional is excessively glassy in the underlying water diffusion, and simulations can vary greatly. In fact, one of our AIMD runs at 300 K shows very little forward hopping over the course of the simulation.

*ii. Correlation Functions*

In order to further compare the hydronium-like hydrated proton structure lifetimes of the different simulation methods, as well as to inspect the effect of eliminating rattling, we calculated the proton identity correlation function and the continuous proton identity correlation function of the filtered and unfiltered trajectories. We first present the proton identity correlation functions, shown in Fig. 5, where the proton identity correlation function is defined as,

$$c(t) = \frac{\langle h(t)h(0) \rangle}{\langle h \rangle} \quad (1)$$

where $h(t)$ is 1 if it is equal to $h(0)$, and 0 if it is not. Comparing the correlation functions of the four simulation methods, we see a more rapid decay in the hydronium-like ion in AIMD at 330 K than AIMD at 300 K. This is consistent with the behavior of more rapid hopping and a higher fraction of multiple hops (which take the excess proton farther away from $h(0)$). We find that EDS-AIMD (OO) and EDS-AIMD (OH) have a decay behavior more similar to AIMD at 330 K than at 300 K, which is not surprising since the EDS bias reduces the water overstructuring of the BLYP functional and hence increases the underlying water diffusion and proton transport. The correlation functions of the filtered trajectories do not differ in spirit from those of the unfiltered trajectories (Fig. S3), but some of the details have been smoothed over.

The more drastic differences are seen in the so-called continuous proton correlation function, defined

$$C(t) = \frac{\langle H(t)H(0)\rangle}{\langle H\rangle} \quad (2)$$

where $H(t)$ is 1 as long as the hydronium-like identity has not changed from that of $H(0)$ and 0 once it has changed. Since rattling will drastically reduce the amount of time before the hydronium identity changes, there is a substantial change in the continuous proton correlation function when one compares the unfiltered and filtered trajectories. First, we compare the continuous proton correlation function of the unfiltered trajectories, shown in Fig. 6. Integrating the correlation functions yield lifetimes of 184 fs for AIMD at 300 K, and 145 fs for EDS-AIMD (OO) at 300 K, 174 fs for EDS-AIMD (OH) at 300 K, 114 fs for AIMD at 330 K. These values are similar to those found by Berkelbach et al.[28] In that paper, the authors note that this is similar to the experimental Eigen-Zundel interconversion time of around 100 fs. We again find EDS-AIMD shows intermediate behavior between AIMD simulations at 300 K and at 330 K.

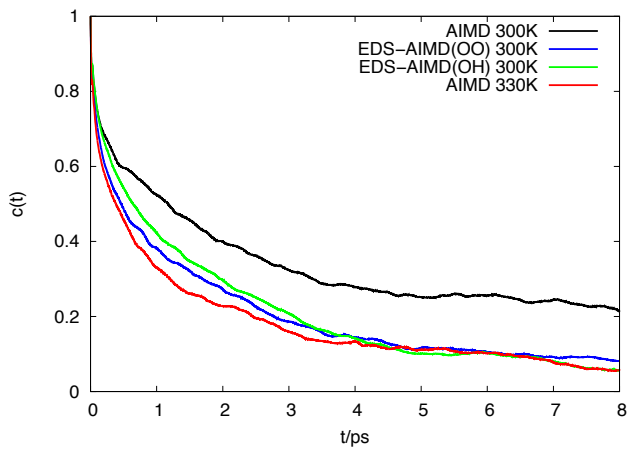

**Figure 5:** Proton correlation function of the unfiltered method using Eq. (1) for the four computational methods studied here.

Calculating the continuous correlation function on the filtered trajectories of course yields very different results. Eliminating rattling removes fleeting hydronium states, and thus extends the apparent lifetime of a particular hydronium. Therefore, the decay of the correlation function is longer, as shown in Fig. 6. We see similar trends as before, where AIMD at 330 K and EDS-AIMD at 300 K decay more rapidly than AIMD at 300 K. Integrating the continuous correlation functions of the filtered trajectories yields lifetimes of 1.69 ps, 1.09 ps, 1.45 ps, and 850 fs for AIMD at 300 K, EDS-AIMD (OO) at 300 K, EDS-AIMD (OH) at 300 K, and AIMD at 330 K, respectively. The authors of Ref. [28] argue this corresponds to the experimental proton hopping time of around 1.5 ps.[9,51] We note that if this interpretation is correct, EDS-AIMD (OH) at 300 K yields results most in line with experiment.

Importantly, the similarity of the integrated lifetimes to well-defined experimentally measured timescales suggest the present analysis method is sensible for separating short- and long-term proton transfer behavior. One would expect that the unfiltered trajectory, which is dominated by rattling events, should reflect the transition time from the Eigen complex to the Zundel complex. Likewise, upon elimination of rattling, the lifetimes should reflect the true hydronium-like structure lifetime – that is, the amount of time before an excess proton truly moves on from one water to another. Given the relative numbers of total hops in the unfiltered and filtered trajectories, and the large difference in the number of hops counted by our analyses from those in Ref. [33], we find it difficult to believe that a correlation function of their "filtered trajectory" would yield something close to what our data presents. Granted, we expect the number of total hops of their unfiltered trajectories to be similar to ours, but find the apparent results of their counting method to be contrary to what we see.

We acknowledge that one potential shortcoming of the analysis presented here, in particular in the analysis of the filtered trajectory, is the overemphasis on forward hopping. Our analysis does count certain back-hops, but the smoothing of rattling events has the clear impact of decreasing multiple hops that include back-hops in favor of mostly counting hops that result in a clear, long-lasting proton transfer event. We argue that in order to assess the part of the proton transfer mechanism which actually results in dislocation of the proton and therefore accounts for the rapid diffusion, it makes sense to separate rattling events. Certainly, rattling is an essential part of the proton transfer mechanism, and previous work[22] has shown hydronium-like core undergoes the so-called special pair dance before ultimately donating the proton.

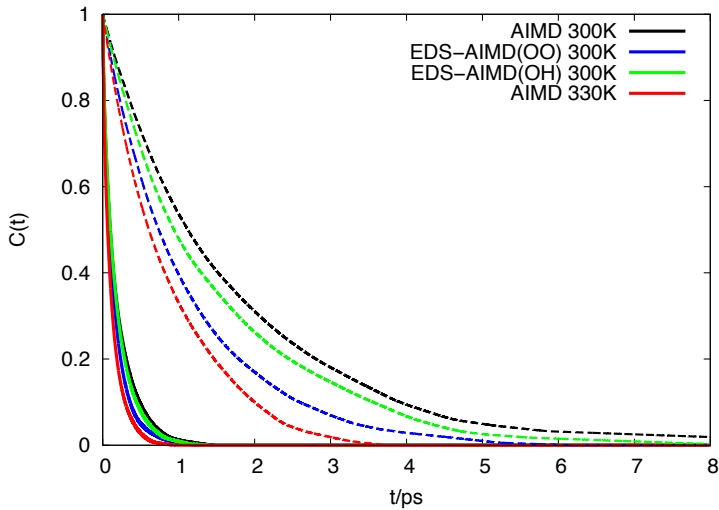

**Figure 6:** The continuous correlation function of the unfiltered (solid lines) and filtered (dashed lines) trajectories for the four methods using Eq. (2).

One crucial feature of being able to pare down rattling events is that it should have no effect on concerted, multiple hopping events. However, simple visual inspection of the forward hopping plots, as well as comparing the number of multiple hopping events of the filtered and unfiltered events, seem to support the notion of the special pair dance. In the unfiltered trajectories, we see the excess proton rapidly oscillating between waters. Not seen in the forward hopping plot, however, is that the identity of the waters often change, i.e., the following motion occurs: B ↔ A ↔ C. That is, water A is rapidly exchanging the proton with water B, then later exchanging the proton with water C. This is exactly the type of behavior described in Ref. [22]. Of course, we still see multiple hopping events, but if the true nature of the majority of these hops were concerted hops, we would expect to see a similar number in the unfiltered and filtered trajectories. As it turns out, there are long periods of the trajectory where little forward activity is occurring, in which the unfiltered trajectory shows constant hopping followed by subsequent back hopping, whereas the filtered trajectory shows no activity at all.

*iii. Multiple Hopping*

In order to quantify the frequency of this type of hopping event (B ↔ A ↔ C), we applied the analysis method in Ref. [31] to the simulation methods employed here; this allows us to quantify the hopping in association with some time scale. To do so, we define a function to quantify the number of water molecules to which the excess proton is associated over a given interval. In a given interval of length $\tau$, the proton is assigned to the water on which it spends the most time during that interval. We then define the hopping function $h_n(\tau)$ as the number of occurrences the proton is associated with $n$ waters over a timespan of $n$ consecutive intervals of length $\tau$. That is, if over a given interval $\tau$, the proton is associated with one water and therefore no hopping happens, $h_1(\tau)$ is incremented; if two intervals are associated with two waters and thus in the case of single hopping, then $h_2(\tau)$ is incremented; if three waters are associated, then $h_3(\tau)$ counts the number of any form of concerted hopping involving three waters including both the forward hopping (A→B→C) and the A↔B↔C phenomenon. The analysis does not consider any hopping details and thus naturally eliminates the rattling and enables a direct focus on the proton hopping occurrences. We can quantify $h_n(\tau)$ for any number $n$, but for the sake of clarity, we have plotted the hopping function for $n = 1 - 3$, shown in Fig. 7.

The decay in $h_1(\tau)$ and $h_2(\tau)$ is a result of the decrease in the number of time segments in which the proton remains on either one or two waters, respectively, as the lengths of the time intervals are increased. We first investigate the effects on $h_1$ as $\tau$ is increased. As $\tau$ is increased, the excess proton is more likely to be associated with a greater number of different waters, and thus the $h_1(\tau)$ curves are all decreasing. Notice that EDS-AIMD (OO) and AIMD at 330 K decays faster than AIMD at 300 K with EDS-AIMD (OH) between AIMD at 300 K and 330K; this indicates more total hops in those methods. The plot for $h_2$ shows that a proton is slightly more likely to be associated with two waters in simulations run with AIMD at 330 K and EDS-AIMD (OO) followed by EDS-AIMD(OH) than in AIMD at 300 K. Consistent with the aforementioned trend that both AIMD at 330 K and EDS-AIMD have more concerting hopping than AIMD at 300 K, the contrast between the methods becomes more significant for $h_3$. At short times, EDS-AIMD (OO) and EDS-AIMD (OH) are similar to AIMD at 330 K, but for longer values of $\tau$, the higher temperature simulation has more concerting hopping, with EDS-AIMD(OH) being the most in agreement with AIMD at 330 K at long times. AIMD at 300 K has fewer $n = 3$ events on all timescales.

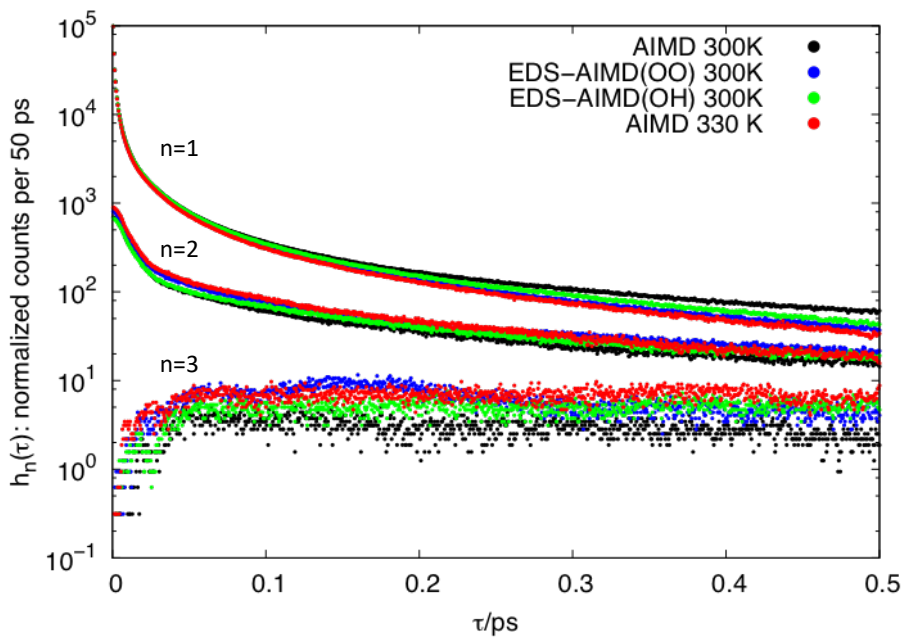

**Fig. 7:** Concerted hopping on time scales between 0 and 0.5 ps.

## B. Diffusion

Table I: Diffusion Coefficients of hydrated excess proton and water from individual runs and their averages. All values are shown in units of Å$^2$/ps. Experimental proton diffusion coefficient comes from ref [52], an water diffusion coefficient comes from ref [53].

| Simulation Method | Replica | $D_{H^+}$ | $\langle D_{H^+} \rangle$ | $D_{H_2O}$ | $\langle D_{H_2O} \rangle$ | $\langle D_{H^+} \rangle / \langle D_{H_2O} \rangle$ |
|---|---|---|---|---|---|---|
| Experiment (300K) | | | 0.94 | | 0.23 | 4.09 |
| AIMD (300K) | 1 | 0.78 | 0.54 ± 0.25 | 0.054 | 0.027 ± 0.027 | 20.0 ± 22.0 |
| | 2 | 0.29 | | 0.00038 | | |
| EDS-AIMD (OO) (300K) | 1 | 0.59 | 0.55 ± 0.04 | 0.13 | 0.13 ± 0.01 | 4.23 ± 0.45 |
| | 2 | 0.51 | | 0.14 | | |
| EDS-AIMD (OH) (300K) | 1 | 0.98 | 0.82 ± 0.16 | 0.14 | 0.15 ± 0.01 | 5.5 ± 1.13 |
| | 2 | 0.66 | | 0.15 | | |
| AIMD (330K) | 1 | 2.07 | 1.65 ± 0.42 | 0.17 | 0.16 ± 0.01 | 10.3 ± 2.70 |
| | 2 | 1.23 | | 0.15 | | |

A principal utility of molecular dynamics is the ability to also capture the dynamical properties of a system in addition to the structural and statistical properties. Clearly, the dynamical property of greatest interest in a system involving hydrated excess protons is the proton self-diffusion constant. The results are summarized in Table I, which also includes the diffusion constants for the individual runs. We specify the self-diffusion constant for individual runs due to the fact that proton diffusion is so dependent on the solvation structure for an individual run (given the overstructuring of the AIMD). For example, compare the self-diffusion constants for the two independent runs of AIMD at 300 K: the calculated self-diffusion constants are 0.78 Å$^2$/ps and 0.29 Å$^2$/ps. That is, in one simulation, the proton dynamics nearly match that of experiment, whereas in the second simulation, the proton diffuses very slowly. We find (in agreement with Ref. [33]) that the proton self-diffusion constant in AIMD at 330 K is higher than the experimental value at 300 K. This is a function of several factors, such as increased thermal energy and decreased proton transfer barrier.

In addition to proton self-diffusion, we also compare the self-diffusion constants of water as seen in Table I. We note that AIMD at 300 K is well below the experimental value of 0.23 Å$^2$/ps.[53] However, there is a clear improvement upon application of the EDS bias,[34, 36] as the water self-diffusion constant increases by an order of magnitude, with EDS-AIMD (OH) closer to experiment. This increase has to do with the reduced effective DFT interactions and therefore reduced stickiness of the water molecules. Increasing the temperature to 330 K also has the pronounced effect of increasing the water diffusion, but not sufficiently to bring it in alignment with experiment.

### C. Radial Distribution Functions

We next examine the radial distribution functions (RDFs) of the hydrated excess proton, which provide a detailed description of the hydration structure of that species. The RDF between hydronium-like oxygen and water oxygen (O*-O) is the most illustrative in describing the solvation structure of the excess proton, so we will examine it first. The data is presented in Fig. 8. First, note that experimental data[54] has a prominent first solvation shell peak centered at 2.45 Å having a height of 4.6. As one moves to longer distances, there is a distinct minimum (valley) at 2.9 Å, and a second solvation shell centered at 4.4 Å having a height of 1.15. None of these aspects are perfectly reproduced by the simulations, but they do reproduce the general shape (and it should be noted that the experiments are carried out at quite high acid concentration). In the four AIMD methods, the position of the first solvation shell peak is shifted to a slightly longer distance; we note all four methods have a first solvation shell peak center at 2.50 Å. AIMD at 300 K has the most pronounced peak, having slightly greater height than EDS-AIMD (OH) at 300 K, EDS-AIMD (OO) at 300 K, and AIMD at 330 K. The larger deviation from experiment comes in the size and position of the second solvation shell peaks. This peak is largely a function of water-water interactions, which are known to be too strong in AIMD, resulting in overstructuring. Interestingly, the second solvation shell in EDS-AIMD (OO) in nearly as overstructured as AIMD at 300 K, despite the fact that the $O_w$-$O_w$ RDF is quite close to experiment, indicating there is some degree of interaction between the hydronium ion and waters in its second solvation shell. We note that EDS-AIMD (OO) was not specifically designed to model the hydronium-water interaction, and the improvement of the water-water solvation structure and water self-diffusion constant more than makes up for the modest effect of the hydrated proton. On the other hand, EDS-AIMD (OH) and AIMD at 330 K are the best in alignment with the second solvation shell of experiment.

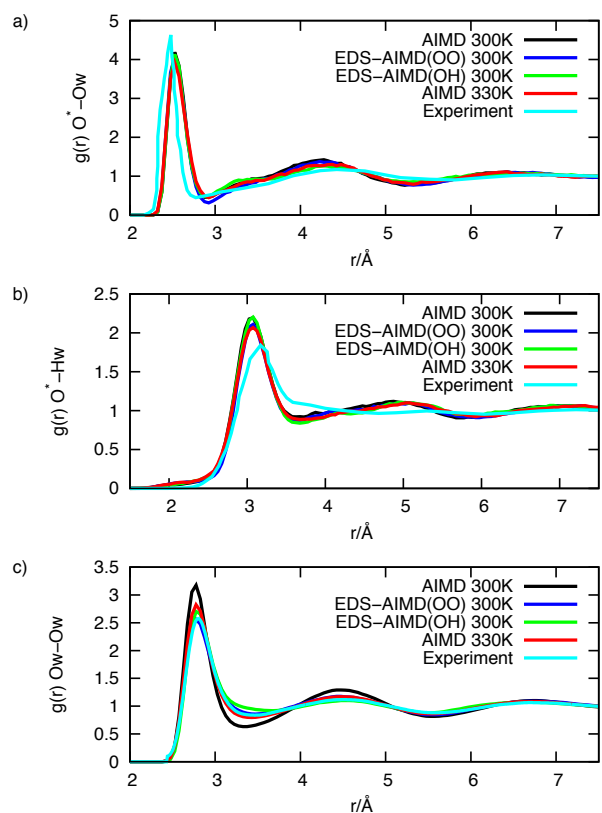

**Figure 8:** Radial distribution functions of (a) O*-Ow, (b) O*-Hw, and (c) Ow-Ow for the four methods studied in this paper in comparison with experiment.

We next compare the O*-H RDF for the four methods. At 300 K, both methods have at least a reasonably accurate first solvation peak center relative to experiment, though both slightly overestimate the peak height. The largest divergence from experiment occurs in the existence of a second solvation shell peak; experiment does not show this peak at all. That is, water hydrogens surrounding a hydronium ion have some structure within the first solvation shell, but none beyond that. Inclusion of quantum nuclear effects into the simulations would mitigate this overstructuring by effectively "delocalizing" the relatively light hydrogen atoms, but that comes at a significant computational cost at the present time. Moreover, as mentioned earlier the experiments are also carried out at high acid concentration.

Also note the existence of the small so-called pre-solation peak centered at around 2.0 Å. It has previously been argued[31] that this peak is beneficial for proton transfer as it readies the hydronium to be able to donate the proton and become a water with 4-fold coordination. In fact, previous work[31] investigating proton transport in AIMD simulations has found that the proton self-diffusion constant is higher in the fragments of trajectories in which pre-solvation is prevalent; conversely, those with effectively no pre-solvation had a relatively stagnant excess proton that underwent few hops. (We remind the reader that AIMD water is notoriously glassy, and thus a given simulation can get "locked in" to a geometry with little to no presolvation.)

To illustrate a advantage of EDS-AIMD relative to standard AIMD, we also present the water $O_w$-$O_w$ RDF in Fig. 8c. We see that the first solvation shell peak of AIMD at 300 K is centered correctly relative to experiment,[55] but is quite a bit larger. Increasing the temperature of the AIMD simulation to 330 K reduces the size of this peak, but does not completely reduce its height to that of experiment. Similarly, AIMD at both temperatures (though much more severely at 300 K) has a too-deep well between the first and second solvation shell, and a second solvation shell peak that is more pronounced than experiment. EDS-AIMD (OO) at 300 K is specifically designed to reproduce the $O_w$-$O_w$ RDF, and we see essentially perfect matching throughout. Additionally, EDS-AIMD (OH) at 300 K is slightly more structured than EDS-AIMD (OO) because it was parameterized to reproduce the classical MB-pol $O_w$-$H_w$ RDF, while EDS-AIMD (OO) was parameterized to reproduce the experimental RDF that incorporated nuclear quantum effects. This has interesting ramifications for the diffusion constant of water, but importantly, this highlights the advantage of using a more accurate AIMD method (EDS-AIMD) rather than simply elevating the temperature of the simulation.

**IV. Conclusions**

Our results lead us to conclude that while concerted excess proton hopping in water does occur, it is not the dominant hopping mechanism. Rather, single proton hopping events occur in much greater number than double or higher order multiple hopping events. We have also simulated with AIMD at both 300 K and 330 K, as well as with an experimentally biased EDS-AIMD at 300 K, and find the dominance of single hopping events in all simulations. While multiple hopping events do occur, most are in fact the result of the proton rattling between several waters, and in most cases the excess proton ultimately returns to the original water, something that was already noted as being unphysically exacerbated in AIMD simulations with overstructured water sixteen years ago.[56] Thus, our current results are consistent with the previously reported proton hopping mechanism involving first a special pair dance.[22]

In order to remove the misleading effect of rattling, we applied a filter to the trajectory, in which we removed all hopping events which are followed directly by a back hop to the original water. This had a large effect on the total number of (apparent) hops, but more interestingly, this greatly reduced the number of multiple hops counted. One would expect that if a multiple hopping event were truly concerted, it would be retained despite the filtering. Filtering the trajectory also has the effect of greatly reducing the number of slingshot events, i.e., where either a back hop is followed by a forward hop. We argue that double hops which are immediately followed by a nullifying back hop are not concerted, and therefore should not be included as a true double hop.

Calculation of the continuous correlation function on the filtered and unfiltered trajectories yield lifetimes similar to experimental values for the proton hopping time and the Eigen-Zundel interconversion time, respectively. In all simulations, AIMD at 330 K decays most rapidly, which is unsurprising given the elevated temperature which facilitates proton hopping.

We find a strong dependence of the multiple hopping rates on temperature, which correlates to a large dependence of the diffusion on temperature. We find that AIMD at 300 K has a proton self-diffusion constant of $0.54 \pm 0.25$ Å$^2$/ps, while AIMD at 330 K has a diffusion constant of $1.65 \pm 0.42$ Å$^2$/ps, about 75 % larger than experiment. This too-rapid diffusion at 330 K relative to experiment would only be further exacerbated after accounting for finite simulation size effects.[57] Thus, caution must be exercised when carrying out AIMD simulations of the excess proton in water (or any aqueous system) at elevated temperatures.

**Supplementary Material**

See the supplementary material for the potential of mean force of the proton transfer coordinate, the forward hopping analysis as a function of time, and the proton correlation function for filtered trajectories.

**Acknowledgments**

This research was supported by the U.S. Department of Energy, Office of Basic Energy Sciences, Separation Science Program of the Division of Chemical Sciences, Geosciences, and Biosciences under Award Number DE-SC0018648. The computational resources for this research were provided by the University of Chicago Research Computing Center.

**Data Availability**

The data that support the findings of this study are available from the corresponding author upon reasonable request.

Supporting Information for

# The Hopping Mechanism of the Hydrated Excess Proton and Its Contribution to Proton Diffusion in Water

Christopher Arnsten,[a] Chen Chen,[b] Paul B. Calio,[c] Chenghan Li,[c] and Gregory A. Voth[c]

[a]Department of Chemistry, Youngstown State University, Youngstown, Ohio, 44505 USA
[b]Department of Mechanical and Nuclear Engineering, Pennsylvania State University, University Park, Pennsylvania, 16802 USA
[c]Department of Chemistry, Chicago Center for Theoretical Chemistry, James Franck Institute, and Institute for Biophysical Dynamics, The University of Chicago, 5735 South Ellis Avenue, Chicago, Illinois 60637, United States

**S1. Proton Transfer Coordinate Potential of Mean Force (PMF)**

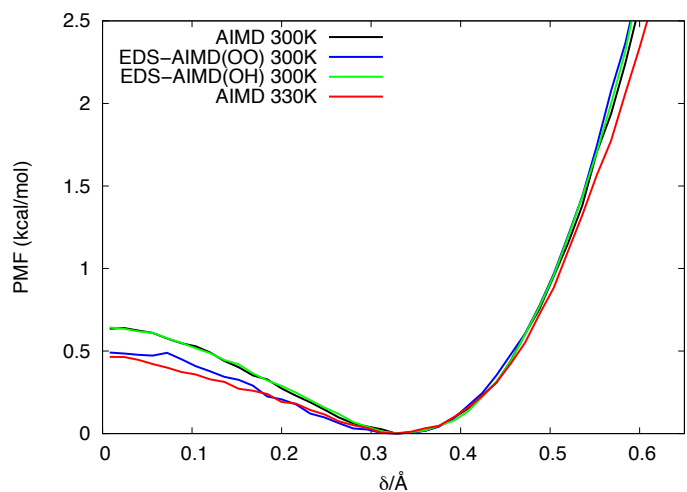

**Figure S1**. Potential of Mean Force for the delta reactive coordinate. In black is AIMD at 300 K, in blue is EDS-AIMD(OO) at 300 K, in green is EDS-AIMD(OH) at 300 K, and in red is AIMD at 330 K.

## S2. Proton Hopping Filtering

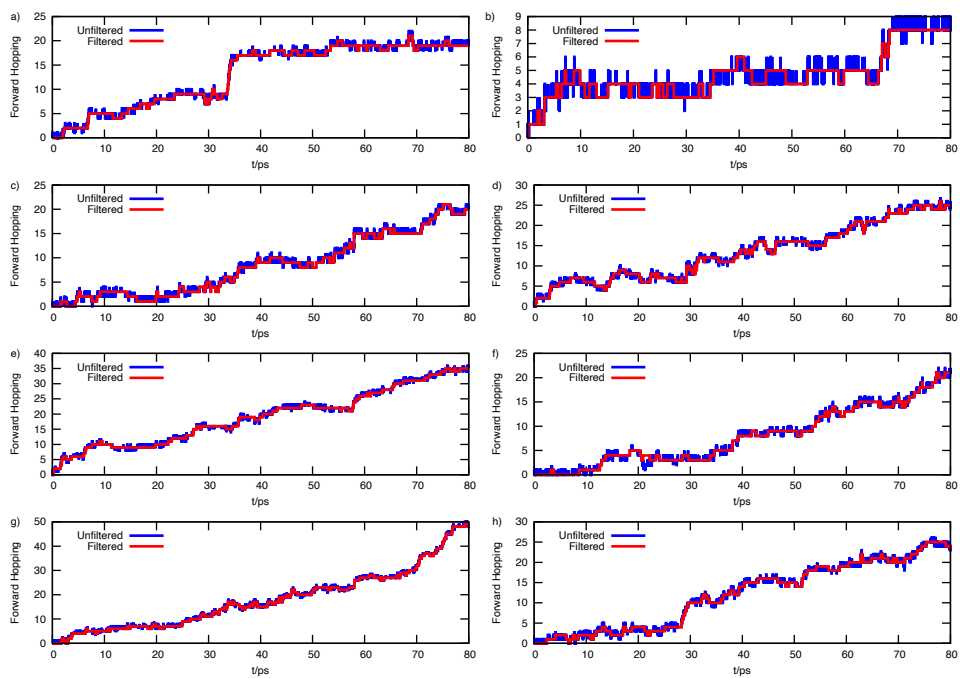

**Figure S2**. Forward hopping as a function of time for the various simulation methods utilized in this work. Fig a-b are AIMD at 300 K, Fig c-d are EDS-AIMD(OO), Fig e-f are EDS-AIMD(OH) at 300 K, and Fig. g-h are AIMD at 330 K at 300 K. In blue is the total forward hopping as a function, of time, and red is the filtered forward hopping.

## S3. Proton Correlation function for Filtered trajectories

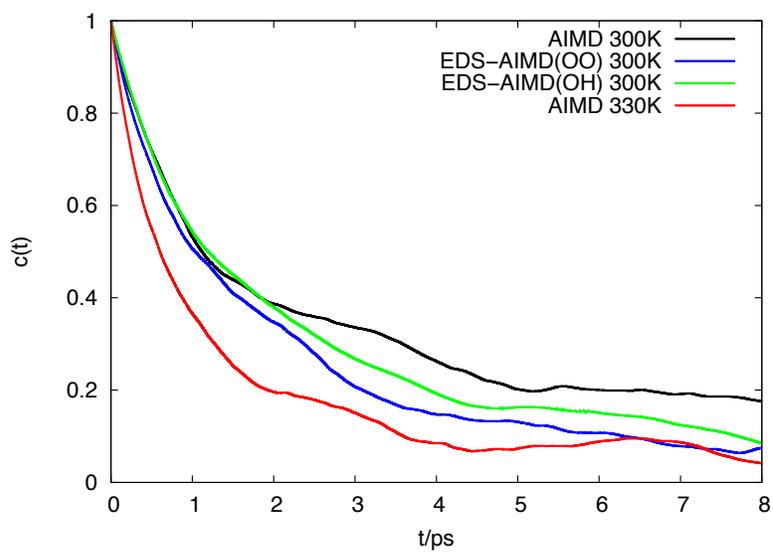

**Figure S3:** Proton correlation function of the filtered method for the four computational methods studied here using Eq. (1) in the main text.